\documentclass{ifacconf}

\usepackage{graphicx,color}
\usepackage{subfigure}
\usepackage{multirow}
\usepackage{amsmath}	
\usepackage{natbib}
\usepackage{amsfonts} 
\usepackage{url}
\renewcommand{\cite}{\citep}

\newcommand{\figref}[1]{Fig.~\ref{#1}}
\renewcommand{\eqref}[1]{Eq.~(\ref{#1})}

\newcommand{\secref}[1]{Section~\ref{#1}}

													{\end{footnotesize}}
													
\usepackage{enumitem}
\newlist{HYP}{enumerate}{1}
\setlist[HYP]{label=\textbf{H\arabic*:}}

\begin{document}

\begin{frontmatter}

\title{FITS: Ensuring Safe and Effective Touchscreen Use in Moving Vehicles} 


\author[First]{Daan M. Pool}
\author[Second]{and Yasemin Vardar}

\address[First]{Control and Simulation, Aerospace Engineering, Delft University of Technology, Delft, The Netherlands (e-mail: d.m.pool@tudelft.nl).}
\address[Second]{Cognitive Robotics, Mechanical Engineering, Delft University of Technology, Delft, The Netherlands (e-mail: y.vardar@tudelft.nl).}


\begin{abstract}
Touch interfaces are replacing physical buttons, dials, and switches in the new generation of cars, aircraft, and vessels. However, vehicle vibrations and accelerations perturb finger movements and cause erroneous touchscreen inputs by users. Furthermore, unlike physical buttons, touchscreens cannot be operated by touch alone and always require users’ visual focus. Hence, despite their numerous benefits, touchscreens are not inherently suited for use in vehicles, which results in an increased risk of accidents. In a recently awarded research project, titled ``\emph{Right Touch Right Time: Future In-vehicle Touchscreens (FITS)}'', we aim to address these problems by developing novel in-vehicle touchscreens that actively predict and correct perturbed finger movements and simulate physical touch interactions with artificial tactile feedback.
\end{abstract}
\begin{keyword} 
touchscreens, vehicle motion, biodynamic feedthrough, tactile feedback, haptics
\end{keyword}

\end{frontmatter}



\section{Introduction}\label{sec:introduction}

In the current vision of future mobility, the next generation of road, air, and water vehicles will be highly automated but with essential high-level functions still controlled by human drivers, pilots, and helmspersons. To enable human users to keep track of vehicle functioning and provide user inputs, touch-based interfaces are seen as the primary means of in-vehicle user interaction \cite{jp:Pitts2012, jp:Stuyven2012, jp:VanZon2020}. This development is driven by touchscreens’ capacity for flexible, direct, and two-way interaction with displayed information \cite{cp:Kaminani2011}, reduced maintenance costs \cite{jp:Orphanides2017}, but foremost due to their near-universal ease-of-use: they provide an interaction mode nearly everyone is highly familiar with from smartphones and tablets \cite{jp:Rogers2005}. 	
Still, entirely carefree use of touch-based interfaces in vehicles has not yet been realized due to two crucial, potentially safety-critical problems:

\begin{enumerate}
    \item \emph{Erroneous touch inputs due to vehicle motion perturbations}. Biodynamic feedthrough (BDFT) of vehicle motions---e.g., turbulence, bumps on the road, waves, sudden turns---causes unintended arm and hand movements and consequently undesired touch inputs \cite{cp:Venrooij2016,jp:Khoshnewiszadeh2021, jp:Cockburn2019, jp:Wang2024}.

    \item \emph{Lack of tactile feedback}. While the grip and feel of traditional buttons and switches provide inherent robustness to motion perturbations \cite{cp:Kaminani2011,jp:Pitts2012}, current touchscreens require drivers, pilots, and helmspersons to always look at a touch interface, taking their eyes `off the road' \cite{jp:Breitschaft2022, cp:Bernard2022}. 
\end{enumerate}

These two factors inevitably cause in-vehicle touchscreens to be pressed `in the wrong place at the wrong time', which is annoying, distracting, or even directly hazardous. For example, driver reaction times have been found to increase five times more when using a touchscreen than when legally drunk \cite{tr:Ramnath2020}.

To overcome these issues, braces to support the hand \cite{jp:Cockburn2019,jp:Schachner2023}, secondary interfaces for use during strong vehicle motions \cite{jp:Large2019}, or basic vibration feedback \cite{jp:Basdogan2020} have been proposed. However, the only way to counter these problems effectively under varying real-vehicle conditions is to 1) actively predict and cancel BDFT-induced touch inputs and 2) use adaptive tactile feedback for restoring the high-fidelity tactile interaction akin to operating physical buttons. A lack of understanding of how varying environmental and biomechanical factors---e.g., motion perturbation characteristics, individual differences in neuromuscular and finger properties, and task-related changes in finger-screen contact---affects BDFT and perceived tactile stimuli prevents augmenting in-vehicle touchscreens with such technologies. 

This paper provides an overview of the recently awarded research project ``\emph{Right Touch Right Time: Future In-vehicle Touchscreens (FITS)}'', funded by the Netherlands Organization for Scientific Research (NWO). In this proposed project, we will perform experiments in motion simulators to systematically disentangle the effects of the varying factors on in-vehicle touchscreen interactions. Using this experiment data, we aim to develop adaptive models of human biodynamics and tactile perception that will enable adaptive BDFT-prediction and tactile feedback technologies on future in-vehicle touchscreens. This paper will present a roadmap of the project's activities (\secref{sec:methods}), as well as some preliminary data for both our BDFT-prediction and tactile feedback work in \secref{sec:results}. 
\section{Methods}
\label{sec:methods}

\subsection{Research focus}

\begin{figure}[hbt]
    \centering
    \includegraphics[width=1\linewidth]{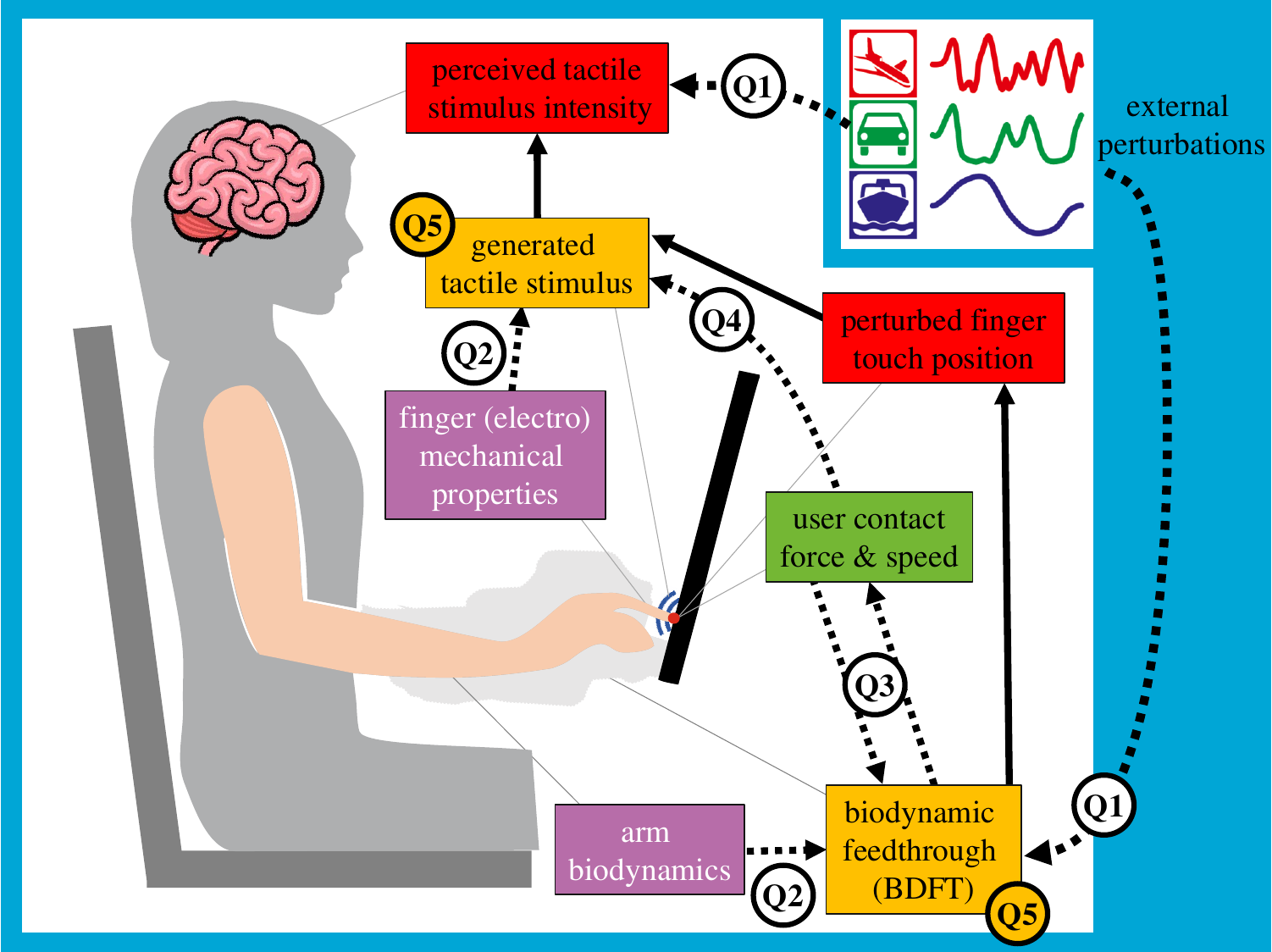}
    \caption{Relations between vehicle motion perturbations, arm biodynamics, finger (electro)mechanical properties, finger contact force and speed, finger touch positions, and perceived tactile feedback. Solid arrows indicate well-established relations; dashed arrows show currently insufficiently understood relations.}
    \label{fig:fits_challenges}
\end{figure}

Implementation of adaptive human-machine interfaces is generally realized using mathematical models that ensure agreeable interface behavior based on measured or predicted system changes. Due to the intricate network of interactions between different factors that affect human-touchscreen interaction in moving vehicles (see \figref{fig:fits_challenges}), models used for BDFT prediction and tactile feedback would need to systematically account for time-varying changes in vehicle motion perturbations, individual differences, and finger contact force and speed to be effective for any user and all environmental conditions. So far, only some isolated parts of the interaction network of \figref{fig:fits_challenges} (solid arrows) are understood and modeled successfully \cite{jp:Vardar2017,jp:Vardar2021,cp:Mobertz2018,jp:Khoshnewiszadeh2021}.
The main focus of our project will be to extend this to the currently insufficiently understood relations, focusing on the three factors explained in the next subsections. 

\subsubsection{Vehicle motion perturbations:}
As indicated with \textbf{Q1} in \figref{fig:fits_challenges}, we will investigate how vehicle motion perturbations affect BDFT and perceived tactile stimuli. So far, tactile feedback for vehicle-related touchscreen operations, e.g., interaction with virtual controls, has been developed and tested in conditions where users interact with devices on a static table. Nonetheless, it is well-known that physical perturbations---e.g., sudden impacts, bumpy roads, waves, and turbulence---directly interfere with perceived tactile sensations due to undesirable ‘masking’ effects \cite{jp:Vardar2018, jp:Vuik2024}. Achieving consistent tactile feedback requires carefully adapting tactile stimuli based on the characteristics of continuously adapting motion perturbations. Similarly, using identified models of BDFT--–i.e., describing how motion perturbations (input) result in unintended finger movements (output)---to correct BDFT-induced inputs is highly effective \cite{cp:Venrooij2016, jp:Khoshnewiszadeh2021}, but so far only for isolated tasks performed in a controlled and stationary laboratory setting. Achieving the same success in a realistic setting, where users will constantly optimize their neuromuscular settings, e.g., by changing muscle tension, under varying motion perturbations, requires a step beyond the current state-of-the-art in BDFT modeling methods \cite{cp:Venrooij2016,jp:Mulder2018}.

\subsubsection{Variation between user physical characteristics:}
Neuromuscular dynamics, e.g., arm/hand/finger inertia and stiffness, and finger (electro)mechanical properties, e.g., skin friction and contact area, vary widely between individuals \cite{jp:Abdouni2017}. As a result, individual differences between users directly impact the effectiveness of any designed BDFT prediction and tactile feedback systems. This implies that ‘one-size-fits-all’ implementations for our proposed touchscreen innovations are unlikely to result in effective and inclusive solutions \cite{cp:Venrooij2016, jp:Khoshnewiszadeh2021}. However, no systematic research on how human neuromuscular dynamics and finger skin properties affect BDFT and tactile perception is available, see \textbf{Q2} in \figref{fig:fits_challenges}, and we lack models that explicitly account for such individual differences.

\subsubsection{Finger contact force and speed:}
Finger contact force and movement speed are crucial, yet insufficiently understood, factors in BDFT and tactile feedback. For example, sustained finger contact with the screen, e.g., dragging gestures, reduces BDFT-susceptibility \cite{jp:Wynne2021} and is lowest when touching a fixed screen location \cite{jp:Khoshnewiszadeh2021}. However, as indicated by \textbf{Q3} in \figref{fig:fits_challenges}, these variations in task-related finger contact force and speed have never been explicitly measured in BDFT experiments to grasp these correlations entirely. The same holds for how, vice versa, BDFT causes changes in screen contact. While recent experiments suggest that BDFT-induced changes in contact force and finger speed \cite{jp:Vuik2024}, this second link between BDFT and screen contact has not yet been explicitly investigated. 

Furthermore, as indicated with \textbf{Q4} in \figref{fig:fits_challenges}, changes in finger contact force and speed affect the generated tactile stimuli, e.g., forces, and thus what is felt by users \cite{jp:Vardar2021,jp:Vuik2024,jp:nam2020} due to induced variations in the finger contact area, stick-slip behavior, and the air gap between the touchscreen and the skin \cite{jp:Vardar2021, jp:serhat2022}. While active tactile stimulus adaptation based on measured finger friction or mechanical impedance has been proposed, these methods break down in the presence of other factors, such as external perturbations. 

\begin{figure*}[t]
    \centering
    \begin{minipage}{0.7\linewidth}
        \centering
        \includegraphics[width=0.74\linewidth]{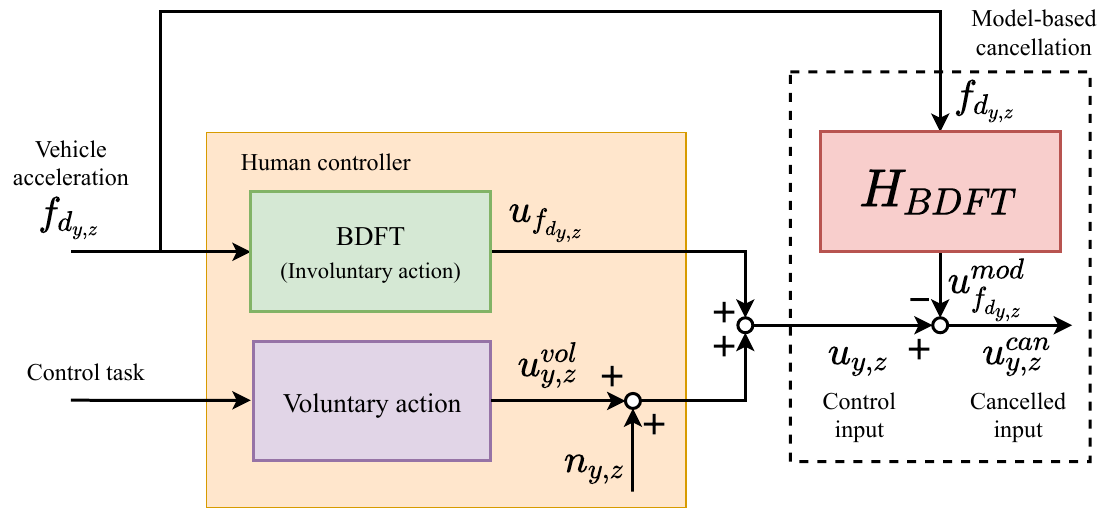}
        \caption{Definition of BDFT prediction and model-based cancellation \cite{jp:Khoshnewiszadeh2021}.}
        \label{fig:bdft_cancellation}
    \end{minipage}
    \hfill
    \begin{minipage}{0.23\linewidth}
        \centering
        \includegraphics[width=0.8\linewidth]{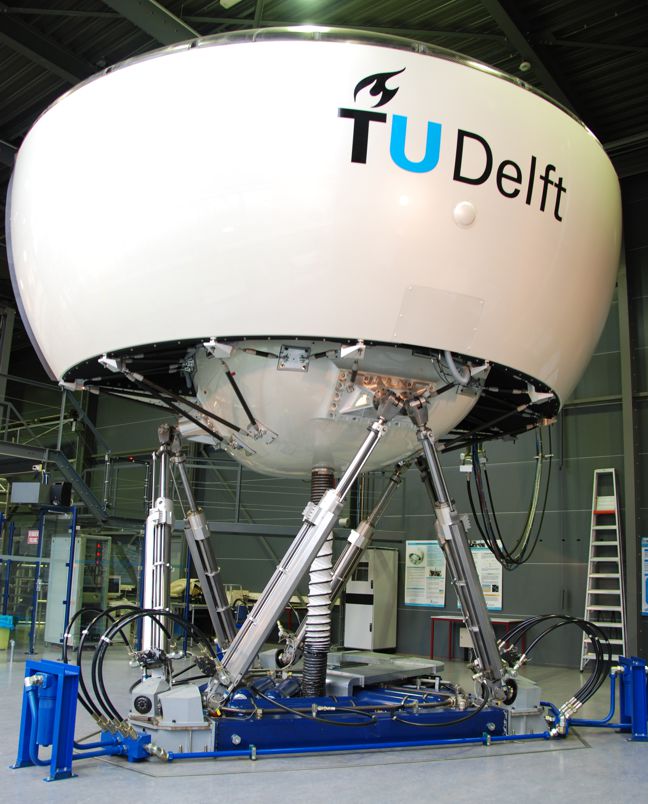}
        \caption{The SIMONA simulator at TU Delft.}
        \label{fig:SIMONA}
    \end{minipage}
\end{figure*}

Overall, we aim to address the open challenges identified in \figref{fig:fits_challenges} by developing models for adaptive BDFT dynamics and adaptive tactile feedback that adjusts itself to biomechanical and task variations. These will be implemented in a Future In-Vehicle Touchscreen (FITS) that will facilitate effective, safe, and comfortable touchscreen operations in future road, air, and water vehicles. More details on the methodology followed for both focus points is provided in the next subsections.


\subsection{Methodology: Biodynamic feedthrough (BDFT)}

\subsubsection{BDFT prediction and cancellation:}

\figref{fig:bdft_cancellation} shows a schematic representation of model-based BDFT cancellation, as defined in \cite{jp:Khoshnewiszadeh2021} and further pursued in this project. This figure shows a 2-dimensional touchscreen input task, for which the finger input coordinates in lateral-horizontal ($y$) and vertical ($z$) screen coordinates are represented by the symbol $u_{y,z}$. For any input task, part of the user's finger input $u_{y,z}$ represents a task-related \emph{voluntary} action, here indicated as $u_{y,z}^{vol}$, which may be executed imperfectly due to human controller nonlinearities and noise, $n_{y,z}$. However, when performed in a moving vehicle, vehicle accelerations and vibrations add a second involuntary BDFT component in $u_{y,z}$. The goal of model-based BDFT cancellation is to, after the fact, remove this contribution of the motion disturbances $f_{d_{y,z}}$ from the user's input $u_{y,z}$. As indicated in \figref{fig:bdft_cancellation}, this can be achieved using a mathematical model of the user's BDFT dynamics, indicated with the red $H_{BDFT}$ block. With a real-time measurement of the vehicle's accelerations $f_{d_{y,z}}$, a BDFT model enables the (potentially real-time) prediction of the BDFT contribution to $u_{y,z}$, here indicated as $u_{{f_d}_{y,z}}^{mod}$. With a predicted BDFT contribution, the real BDFT $u_{{f_d}_{y,z}}$ can be removed by calculating a `cancelled' input $u_{y,z}^{can}$, which with an accurate BDFT model $H_{BDFT}$ will approximate $u_{y,z}^{vol}$. We focus on sufficiently accurate and adaptive BDFT models for effective model-based BDFT cancellation under varying real-world vehicle conditions. Furthermore, as in many cases touchscreen users will be affected by BDFT also before their fingers are in contact with the screen, BDFT cancellation will be developed for two cases, where users' fingers are 1) in sustained contact with the screen (i.e., dragging) and 2) are being moved towards the screen (i.e., button press). 

\subsubsection{Motion perturbation signals:}

To identify adaptive BDFT models, several experiments will be performed in TU Delft's SIMONA Research Simulator, see \figref{fig:SIMONA}. To facilitate the use of well-known spectral system identification methods \cite{cp:VanPaassen1998}, multi-sine motion perturbations as defined by \eqref{eq:multisine}, as also done in \cite{cp:Mobertz2018}, will be used for all experiments:
\begin{equation}
f_{d}(t)=\sum_{k=1}^{N_{d}} A_{d}[k] \sin \left(\omega_{d}[k] t+\phi_{d}[k]\right)
\label{eq:multisine}
\end{equation}
In \eqref{eq:multisine}, $A_d[k]$, $\omega_d[k]$, and $\phi_d[k]$ represent the amplitude, frequency and phase shift of each sine in the multi-sine motion perturbation signal, respectively. To capture the crucial differences in BDFT experienced in different vehicles---e.g., due to vertical turbulence accelerations in aircraft, sustained lateral (steering) and forward (accelerating/braking) accelerations in cars, and low-frequency vertical and rolling/pitching motion in ships---a realistic set of motion perturbations representative for road, air, and water vehicles will be designed for our simulator experiments. For this, we will approximate real-world vehicle motions with multi-sine signals, as also successfully done in \cite{cp:Kolff2019}.

\subsubsection{BDFT identification and modeling:}

\figref{fig:bdft_cancellation} shows that the involuntary BDFT-related component in a touchscreen input $u_{y,z}$ can be considered as a dynamic response to the experienced vehicle accelerations. When multi-sine motion disturbance signals $f_d$ are used, this allows for direct identification of a BDFT frequency response at all excitation frequencies of $f_d$, according to:
\begin{equation}
    \hat{H}_{BDFT}(j\omega_d) = \frac{S_{f_{d},u_{y,z}}(j\omega_d)}{S_{f_{d_{y,z},}f_{d_{y,z}}}(j\omega_d)}  
    \label{eq:bdft_ident}
\end{equation}
\noindent where the cross power spectral density $S_{f_{d},u_{y,z}}$ and power spectral density $S_{f_{d_{y,z},}f_{d_{y,z}}}$ are calculated directly from measured $u_{y,z}$ and $f_d$ signals \cite{cp:VanPaassen1998}. \eqref{eq:bdft_ident} shows that for multi-sine $f_d$ signals and with high signal-to-noise ratios, all power at $\omega_d$ in $u_{y,z}$ can be directly attributed to BDFT. 

While the identification of $\hat{H}_{BDFT}(j\omega_d)$ will facilitate grasping the wide variation in expected BDFT dynamics, the ultimate goal is to identify parameterized BDFT models \cite{cp:Venrooij2016,cp:Mobertz2018}, as defined in \eqref{eq:bdft_model}:
\begin{equation}
    H_{BDFT}(s) = \frac{G_{BDFT} \omega_{BDFT}^2}{ s^2 + 2\zeta_{BDFT}\omega_{BDFT} s + \omega_{BDFT}^2}
    \label{eq:bdft_model}
\end{equation}
As shown in \figref{fig:bdft_cancellation}, an accurate $H_{BDFT}$ model---i.e., with an accurate estimate of the BDFT model's gain $G_{BDFT}$, natural frequency $\omega_{BDFT}$, and damping ratio $\zeta_{BDFT}$ parameters---will enable effective model-based BDFT cancellation. In our project, we focus on investigating how the three parameters of this model need to be adapted -- e.g., using a linear parameter-varying BDFT model -- to counter BDFT for all expected users in moving air, road, and water vehicles. 

\subsection{Methodology: Tactile feedback}

\subsubsection{Feedback technology:} We use electrostatic actuation (i.e., electrovibration or electroadhesion) as the tactile feedback technology because it can create perceivable tactile stimuli without mechanical motion and is independent of touchscreen size \cite{jp:Basdogan2020}, making electrostatic displays a good candidate for future vehicle cockpits. These devices display an electrostatic force that is generated by applying an alternating voltage signal to the conductive layer of a capacitive touchscreen. When users slide their fingers on the display, this alternating force, which pulls the finger skin towards the screen, is perceived as the tactile stimulus. By controlling the input voltage signal, various tactile controls relevant to touchscreen interactions, such as sliders, buttons, and knobs, can be rendered. 

\subsubsection{Modelling finger-touchscreen contact dynamics:} 
\begin{figure}[h]
    \centering
        \includegraphics[width=0.9\linewidth]{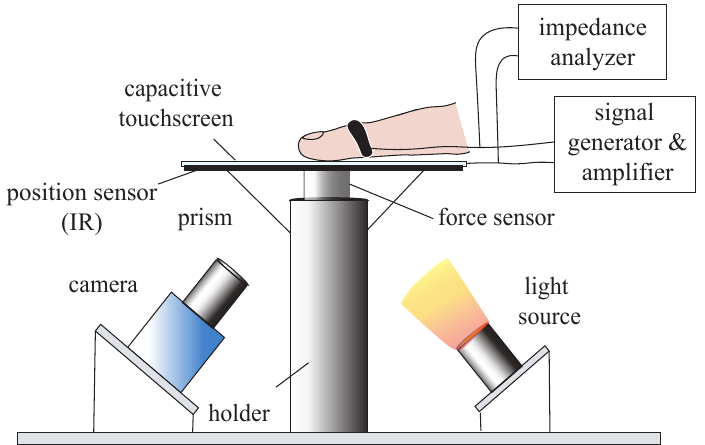}
       \caption{Illustration of the experimental apparatus for measuring physical changes at finger-display contact. Adapted from \cite{jp:serhat2022}.}
        \label{fig:apparatus}
\end{figure}

The tactile stimuli generated by electrostatic displays depend on finger skin properties and changes in finger contact: the same applied voltage signal can generate different electrostatic forces depending on these conditions \cite{jp:Vardar2017, jp:Vardar2021}. We aim to identify how systematic changes in these varying conditions affect the generated electrostatic force and resultant tactile perception. For this goal, we use an experimental apparatus that can monitor physical changes at the contact---e.g., high-resolution finger contact area deformations, finger contact forces, and electrical impedance and current---under varying conditions: individuals with different finger biomechanical properties and applied force and speed. An illustration of an early prototype of this apparatus is shown in Fig.~\ref{fig:apparatus} and in \cite{jp:serhat2022, jp:nam2020}. By conducting systematic measurements using this apparatus, we can identify relevant factors and derive mathematical models representing the underlying interactions. For the modelling, a nonlinear JKR contact model will be adapted for electrostatic pressure \cite{jp:basdogan2020model} and varying conditions affecting finger deformation. 

\subsubsection{Adaptive tactile feedback:} Besides modeling the finger-display contact, we conduct psychophysical experiments in motion simulators, see Fig.~\ref{fig:SIMONA}, to measure the detection thresholds of human participants for tactile stimuli, i.e., tactile controls, in the absence and presence of simulated vehicle motion perturbations; see \cite{jp:Vuik2024} for more details on the experimental methodology. Our goal is to carefully disentangle the physical effects from the perceptual effects. Therefore, using the obtained finger-display contact model, the applied input voltage to the display can be adapted for varying finger properties of the users, task-related changes in finger-display contact, and involuntary changes due to vehicle motion perturbations. The threshold differences under these conditions in the absence and presence of vehicle perturbations will reveal perceptual masking effects. Then, we will model the influence of perceptual masking and integrate it within the tactile feedback controller.   

\section{Preliminary Results and Discussion}
\label{sec:results}

\subsection{Biodynamic feedthrough}

\begin{figure}[b]
   \centering
   \subfigure[Finger trajectory]{\includegraphics[width=0.8\linewidth]{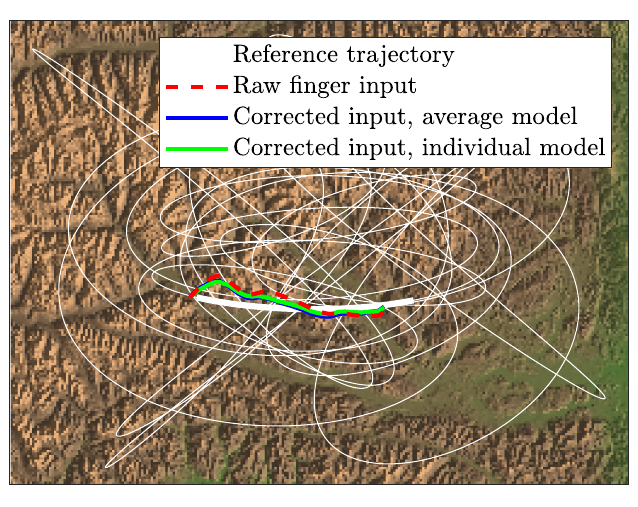}%
                           \label{sfig:bdft_example_trajectory}}\\
   \vspace{-0.4cm}%
   \subfigure[BDFT dynamics]{\includegraphics[width=0.9\linewidth]{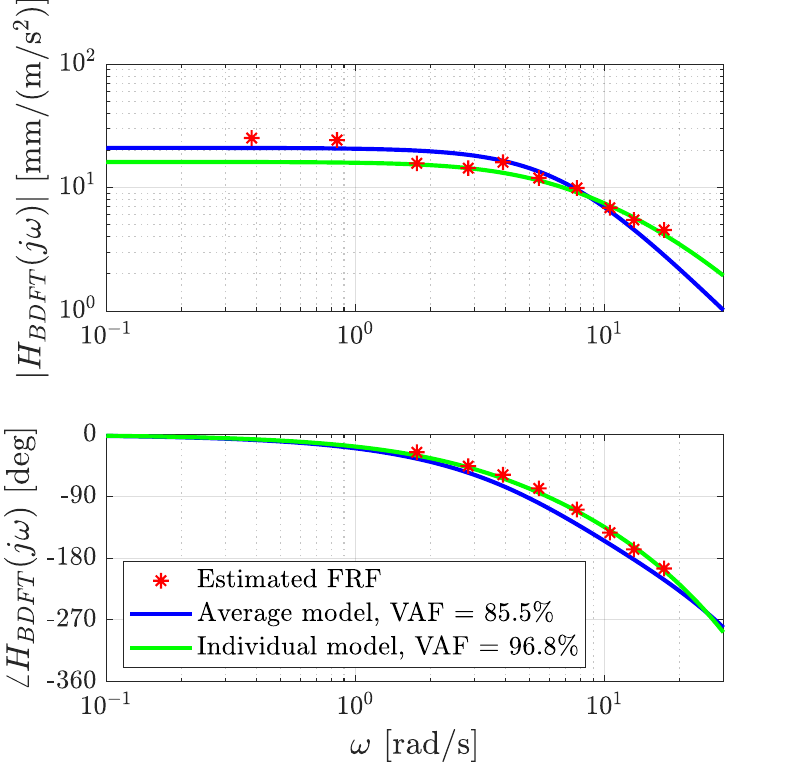}%
                           \label{sfig:bdft_example_bode}}
   \caption{Example BDFT modeling results for an individual participant from \cite{jp:Khoshnewiszadeh2021}.}
   \label{fig:bdft_results}
\end{figure}

\figref{fig:bdft_results} shows example BDFT modeling and cancellation results from the previous experiment of \cite{jp:Khoshnewiszadeh2021}. \figref{sfig:bdft_example_trajectory} shows the commanded finger trajectory in the a continuous dragging task with the white line, where the thicker portion of the trajectory shows a 1-second highlight. For this same 1-second period, the red line shows the measured finger input trajectory $u_{y,z}$ under vertical turbulence. Examples of the `cancelled' input $u_{y,z}^{can}$ obtained with a BDFT model that represents the average BDFT dynamics across all participants (blue line) or the individual users' BDFT dynamics (green line) are also shown. Both cancelled input trajectories show better reference trajectory following, due to successful removal of part of the BDFT component in $u_{y,z}$.
\figref{sfig:bdft_example_bode} shows a Bode plot of the corresponding identified BDFT models, where again the average and individual results are indicated in blue and green, respectively. The estimated frequency response $\hat{H}_{BDFT}(j\omega_d)$ at the 10 $\omega_d$ frequencies of \cite{jp:Khoshnewiszadeh2021} is shown with red markers. \figref{fig:bdft_results} provides a first indication of the importance of matching BDFT models to individual users: for this example the quality-of-fit---here expressed in terms of the Variance Accounted For (VAF)---increases from 86\% (average model) to 97\% for an individual BDFT model. As can be seen in \figref{sfig:bdft_example_trajectory}, this improved model accuracy directly enables superior BDFT cancellation.

With upcoming experiments, we expect to explicitly identify the adaptive mechanisms in touchscreen BDFT and develop explicit scheduling of key BDFT parameters to match individual users' adaptations to varying motion perturbations and task demands. Furthermore, to validate the `lumped' models of BDFT dynamics according to \eqref{eq:bdft_model} we expect to use for implementation on future in-vehicle touchscreens, we will also use more high-fidelity BDFT models and novel measurement techniques---i.e., using external cameras and the OpenPose library\footnote{\url{https://viso.ai/deep-learning/openpose/}}---to relate our `lumped' models to the true BDFT-induced arm, hand, and finger movements. 

\subsection{Tactile feedback}
\begin{figure}[hbt]
    \centering
        \includegraphics[width=1\linewidth]{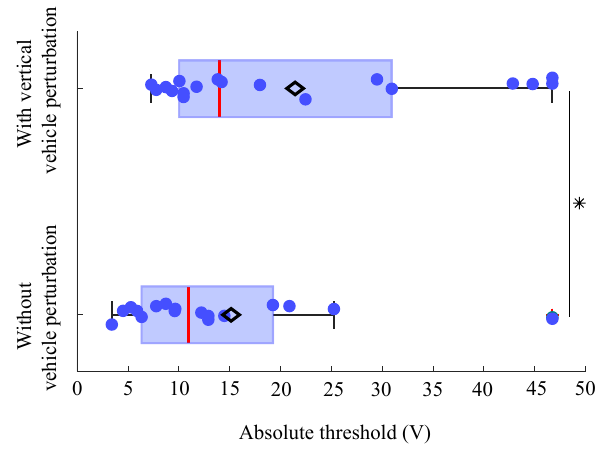}
       \caption{Example detection thresholds measured in the absence and presence of vehicle perturbations, adapted from \cite{jp:Vuik2024}.}
        \label{fig:threshold}
\end{figure}

\figref{fig:threshold} visualizes the results of an example threshold experiment for detecting a 0.2-second electrostatic tactile stimulus---i.e., edge of a virtual button---generated by applying 125~Hz voltage input to a capacitive touchscreen. The experiments were conducted in the SIMONA simulator (see \figref{fig:SIMONA}) with 18 participants in the absence and presence of vehicle perturbations; see \cite{jp:Vuik2024} for details. In \figref{fig:threshold}, blue dots represent the thresholds of each participant, and the diamonds, $\Diamond$, indicate the means. 

In these experiments, we did not adapt the tactile feedback to any varying factors indicated in the previous sections. A Bonferroni-corrected Wilcoxon signed rank test revealed that the thresholds are significantly higher during vehicle perturbations (p $< $0.05). Also, large variances between the thresholds of different participants are visible. These preliminary experiments underscore the importance of adapting tactile feedback to individual users and the experienced vehicle motion perturbations. 

\section{Conclusions and Outlook}
\label{sec:conclusion}

This paper summarizes the goal, key objectives, methodologies, and some preliminary results of the recently awarded FITS project, that aims to develop `Future In-vehicle Touch Screens' that are much better suited for use in moving vehicles than currently available touch interface technology. This will be achieved by 1) developing adaptive biodynamic feedthrough (BDFT) models that can be used to detect and counter erroneous touch inputs that occur due to physical vehicle motion accelerations and 2) by using artificial tactile feedback technology to restore a true sense of touch to in-vehicle touchscreens. In the FITS project, simulator experiments with realistic vehicle motion perturbations will be used to systematically disentangle the effects of crucial varying factors---i.e., individual differences in physical characteristics, motion perturbations, and task-related finger contact force and movement speed---on BDFT and the perception of tactile feedback stimuli. From this experiment data, practical adaptive models of human biodynamics and tactile interactions will be extracted, and used to enable effective BDFT-prediction and tactile feedback in realistic, varying, real-world vehicle settings. The involvement of industry partners from across the automotive, aviation, and maritime domains will enable us to, for the first time, actively account for the broad range of motion disturbances that touchscreen users may encounter and capture their effects on the dynamics of touchscreen interaction. 
  
\section*{Acknowledgments}

This publication is part of the project with number 20624 of the Open Technology Program financed by the Netherlands Organization for Scientific Research (NWO).


\bibliography{bib/references}



\end{document}